\documentstyle[12pt]{article}
\newcommand{\nonum}{\nonumber}
\newcommand{\nn}{\nonumber \\}
\newcommand{\ri}{\right}
\newcommand{\lf}{\left}
\newcommand{\del}{\partial}
\newcommand{\cob}{\delta}    
\newcommand{\al}{\alpha}

\newcommand{\bt}{\beta}

\newcommand{\ep}{\epsilon}
\newcommand{\vep}{\varepsilon}
\newcommand{\th}{\theta}
\newcommand{\Ga}{\Gamma}
\newcommand{\ga}{\gamma}
\newcommand{\Si}{\Sigma}
\newcommand{\si}{\sigma}
\newcommand{\ka}{\kappa}

\newcommand{\la}{\lambda}

\newcommand{\riya}{\rightarrow}

\newcommand{\lrya}{\leftrightarrow}

\newcommand{\bl}{\Bigl}
\newcommand{\br}{\Bigr}
\newcommand{\Bl}{\Biggl}
\newcommand{\Br}{\Biggr}

\newcommand{\wg}{\wedge}
\newcommand{\tr}{{\rm{tr}}}
\newcommand{\Tr}{{\rm Tr}}

\renewcommand{\^}{\hat}
\newcommand{\half}{{1 \over 2}}
\newcommand{\<}{\langle}
\renewcommand{\>}{\rangle}

\renewcommand{\b}[1]{\bar{#1}}
\renewcommand{\t}[1]{\tilde{#1}}
\newcommand{\tg}{\tilde{g}}
\renewcommand{\o}{\over}
\newcommand{\Ts}{{\rm T}^{\ast}}
\newcommand{\T}{{\rm T}}
\newcommand{\bfb}{{\bf b}}
\newcommand{\bfc}{{\bf c}}
\newcommand{\bfB}{{\bf B}}
\renewcommand{\d}{{\rm d}}
\newcommand{\Ws}{\not\!\! W}

\begin{document}

\vskip 0.5 truecm
{\baselineskip=14pt
 \rightline{
 \vbox{
       \hbox{UT-787}
       \hbox{September  1997}
}}}
~~\vskip 5mm

\vfill

\begin{center}
{\Huge  SO(9,1) invariant matrix formulation of supermembrane }
\end{center}
\vskip .5 truecm
\centerline{ Kazuo Fujikawa and Kazumi Okuyama}
\vskip .4 truecm
\centerline {\it Department of Physics,University of Tokyo}
\centerline {\it Bunkyo-ku,Tokyo 113,Japan}
\vskip 0.4 truecm
\centerline{\sf fujikawa@hep-th.phys.s.u-tokyo.ac.jp
 , okuyama@hep-th.phys.s.u-tokyo.ac.jp}

\makeatletter
\@addtoreset{equation}{section}
\def\theequation{\thesection.\arabic{equation}}
\makeatother

\vfill
\vskip 0.5 truecm
\baselineskip 6.5mm
\centerline{\bf Abstract}
\vskip 3mm
An $SO(9,1)$ invariant  formulation of an 11-dimensional supermembrane
is presented by combining an $SO(10,1)$ invariant treatment of
reparametrization symmetry with an $SO(9,1)$ invariant $\theta_{R} = 0$ 
gauge of $\kappa$-symmetry. The Lagrangian thus defined consists of 
polynomials in dynamical variables (up to quartic terms in $X^{\mu}$ and 
up to the eighth power in $\theta$), and  reparametrization BRST symmetry
is manifest. The area preserving diffeomorphism is consistently 
incorporated and the area preserving gauge symmetry is made explicit. 
The $SO(9,1)$ invariant  theory contains terms which cannot be induced by 
a naive dimensional reduction of higher dimensional supersymmetric 
Yang-Mills theory. The $SO(9,1)$ invariant Hamiltonian and the generator 
of area preserving diffeomorphism together with the supercharge are
matrix regularized by applying the standard procedure.  As an application 
of the present formulation, we evaluate the possible central charges in 
superalgebra both in path integral and in canonical (Dirac) formalism, 
and we find only the two-from charge $[ X^{\mu}, X^{\nu}]$.

\vfill

\newpage

\section{Introduction}
A matrix formulation \cite{dWHN,dWLN} of an 11-dimensional 
supermembrane \cite{bst}-\cite{rev-mem} received 
much  attention recently in connection with a possible non-perturbative 
analysis of the  so-called M-theory \cite{schwarz,duff, bfss}. 
The matrix formulation so far is based on the 
light-cone gauge formulation \cite{tanii}, which simplifies 
much the structure of
the action. Recently, we presented  a Lorentz covariant 
matrix formulation of a bosonic membrane\cite{fuji-ko1}, by extending a
full covariant
BRST  formulation of the  bosonic membrane in Ref.\cite{fuji-kubo1}. In
the present  paper, we 
present an $SO(9,1)$ invariant formulation of the supermembrane by 
combining the  manifestly Lorentz covariant treatment of  reparametrization
symmetry with  the  $SO(9,1)$ invariant  $\theta_{R} =0$ gauge of 
$\kappa$-symmetry, which has  been proposed
recently\cite{APS,berg-kall,kall}. 
The theory thus
formulated consists of finite polynomials in dynamical variables , and
reparametrization BRST symmetry is explicit.  The area preserving
diffeomorphism and the area preserving gauge
symmetry are also made manifest. The matrix formulation is obtained by
applying
the standard procedure. 
This formulation, which preserves most  of the Lorentz boost
symmetry,   inherits much of the structure of the original supermembrane,
and  for example, it contains terms which cannot be obtained by a naive 
dimensional  reduction of higher dimensional supersymmetric Yang-Mills
theory. Another characteristic of the present formulation is that the
$\theta_{R}=0$ 
gauge becomes  singular for a naive ``double dimensional reduction'' of the
supermembrane to the Type IIA string; if such a reduction exists , it
should 
be non-perturbative one in the present formulation. 
As an application of this formulation, we examine the possible central
charges in superalgebra in path integral as well as in canonical (Dirac)
formulations.
We find only the (possible) two-form central charges but no five-form
charges,  as is expected for a supermembrane.
We emphasize that the light-cone formulation \cite{dWHN} and the present 
$SO(9,1)$ invariant formulation, though Lagrangians have quite different 
appearance, in fact describe an identical theory in the domain where both 
of the gauge conditions are well-defined.    

To make this paper self-contained , we here recapitulate the basic
definition of the supermembrane\cite{bst}\cite{rev-mem}: The action
consists of two terms, the
Dirac term and the Wess-Zumino term. The Dirac term $S_D$ is written as 
\begin{eqnarray}
S_D &=&\int_W\d^3\si ~\half \sqrt{-g}(1-g^{ab}h_{ab}) \nonum \\
&=& \int_W\d^3\si ~\half (-\det \tilde{g}-\tilde{g}^{ab}h_{ab})
\label{Dirac term}
\end{eqnarray}
where $\tilde{g}^{ab}=\sqrt{-g}g^{ab}$,  and $\si^a~(a=0,1,2)$ are 
membrane 
world-volume coordinates. The variables  $h_{ab}$ are  induced metric on
the 
membrane world-volume $W$
\begin{equation}
h_{ab}=\eta_{\mu\nu}\Pi^{\mu}_a\Pi^{\nu}_b
\end{equation} 
where  the flat $D=11$ target space-time metric is defined by
$\eta_{\mu\nu}= 
{\rm diag}(-1, +1,\cdots, +1)$, and 
\begin{equation}
\Pi_a^\mu =\del_a X^\mu - i\b{\th}\Ga^\mu \del_a \th
\end{equation}
with a 32-component Majorana spinor $\th$ and $\{ \Gamma^{\mu},
\Gamma^{\nu}\}
= 2\eta^{\mu\nu}$.

The Wess-Zumino term is written as 
\begin{equation}
 S_{WZ}=-\int_W \half a_3
\label{Wess-Zumino}
\end{equation}
with a 3-form $a_3$ 
\begin{equation}
a_3=\Sigma_{\mu\nu}\al^{\mu\nu}_2
=\Sigma_{\mu\nu} \lf[\Pi^{\mu}\Pi^{\nu}-\Sigma^{\mu}
( \Pi^{\nu}-{1 \over 3}\Sigma^{\nu}) \ri]
\end{equation}
The 3-form  $a_3$ is a potential of a closed 4-form $h_4$
\begin{equation}
h_4=\d a_3 = \d\Sigma_{\mu\nu}\Pi^{\mu}\Pi^{\nu}
=  -i \d\bar{\th}\Gamma_{\mu\nu}\d\th \Pi^{\mu}\Pi^{\nu}
\end{equation}
In this paper we often use the form notation, which simplifies many of the
equations. We here defined the basic 1-forms by
\begin{equation}
\Sigma_{\mu ..\nu} = i \d\bar{\th}\Gamma_{\mu ..\nu}\th
\end{equation}
with the statistics convention  $\d\th=\del_a\th \d\si^a=\d\si^a\del_a\th$,
and 
\begin{equation}
\Pi^{\mu} =  \Pi_a^\mu \d\si^a=\d X^{\mu}+\Si^{\mu}
\end{equation}
Our definition of exterior derivative is the standard one on the {\em
bosonic}
manifold,i.e.,
\begin{equation}
\d(A_p\wg B_q)=\d A_p\wg B_q+(-1)^pA_p\wg\d B_q
\end{equation}
for $p$-form $A_p$ and $q$-form $B_q$.

The supermembrane action has several symmetries : reparametrization
symmetry, which is manifest in the action, and global SUSY and local 
$\ka$-symmetries. 
The target space global SUSY is defined by  
\begin{eqnarray}
&&\cob_{SUSY}X^{\mu}=i\bar{\ep}\Gamma^{\mu}\th \equiv l^{\mu}\nonum \\
&&\cob_{SUSY}\th = \ep
\end{eqnarray}
and $\Pi^{\mu}$ is invariant, $ \cob_{SUSY}\Pi^{\mu} =0$.
The $\ka$-symmetry is defined by 
\begin{eqnarray}
\cob_{\ka}\th &=& (1+\ga)\ka \nonum\\
\cob_{\ka}X^{\mu} &=& i\bar{\th}\Ga^{\mu}\cob_{\ka}\th \nonum\\
\cob_{\ka} \tilde{g}^{ab} &=& -K^a_c  \tilde{g}^{cb}
\label{kappa-trans}
\end{eqnarray}
where $\ga$, which is a world-volume analogue of $\gamma_{5}$,  is defined
by 
\begin{equation}
\ga = {1 \over 3!}\ep^{abc}\ga_a\ga_b\ga_c
\label{gamma}
\end{equation}
in terms of  the induced $\ga$-matrices
\begin{equation}
\ga_a=\Pi^{\mu}_a\Ga_{\mu} \hspace{3mm},
\hspace{3mm}\{\ga_a,\ga_b\}=2h_{ab}
\end{equation}
The convention of the anti-symmetric tensor is 
\begin{equation}
\vep_{012}=1 \hspace{3mm}, \hspace{3mm} \vep^{012}=-1
\end{equation}
and 
\begin{equation}
\ep_{abc}=\sqrt{-g}\vep_{abc} \hspace{3mm}, \hspace{3mm}
\ep^{abc}={1\over\sqrt{-g}}\vep^{abc}
\end{equation}
The coefficient $K$ in (\ref{kappa-trans}) is given by  
\begin{eqnarray}
K^a_b &=& i\ep^{acd}\del_b\bar{\th}\ga_{cd}\cob_{\ka}\th +(a \lrya b)
\nonum \\
 && -{2i \over 3}(\del_e \bar{\th} \ga^e \ka) [f(1,1)+f(1,\bar{h})
+f(\bar{h},\bar{h})]^a_b
\end{eqnarray}
where
\begin{equation}
f(A,B)=\tr A \tr B -\tr AB +AB+BA-A\tr B -B\tr A 
\end{equation}
and $\b{h}^a_b=g^{ac}h_{cb}$.We raise and lower the world-volume indices
$a,b$
by the metric $g_{ab}$. We note that 
\begin{equation}
\ga^2= \det \bar{h}={1\o 3!}\tr\bl[\b{h}f(\b{h},\b{h})\br]
\end{equation}
before the use of equations of motion for $g_{ab}$.

\section{$SO(9,1)$ invariant gauge fixing}
 
In this section we define an $SO(9,1)$ invariant gauge for the
supermembrane.
The 32-component Majorana spinor is an irreducible representation of 
$SO(10,1)$, and any algebraic gauge fixing of $\kappa$-symmetry generally
breaks the full
 $SO(10,1)$ symmetry. The basic idea of $\theta_{R}\equiv 
\half(1-\Ga_{11})\th=0$ gauge is to
decompose 
\begin{equation}
\begin{array}{rcl}
\th &=& \th_L +\th_R \\
{{\bf 32}}&=& {{\bf 16}}_L \oplus {{\bf 16}}_R \\
SO(10,1) &\supset & SO(9,1)
\end{array}
\end{equation}
We also decompose $\{X^{\mu}\}$ as $X^{\mu}=(X^m,X^{11})$,where we use 
$\mu$ for a  11-dimensional  index and $m$ for a  10-dimensional 
index.Note that the eleventh element $\Ga_{11}$ of 
$D=11$ $\Ga$-matrices
$\{\Ga^{\mu}\}$ is identified with $D=10$ chirality matrix 
$\Ga_0\cdots\Ga_9$.
The $\th_R =0$ gauge fixing of $\ka$-symmetry \cite{APS,berg-kall,kall} is
not $D=11$ covariant but it is $D=10$ covariant. This gauge preserves most
of the Lorentz
boost symmetry compared to  the light-cone gauge, which is based on  
$SO(10,1) \supset SO(1,1)\times  SO(9)$.

In the following, we put $\hat{}$ on 
the objects on the world-volume which consist of 10-dimensional variables.
For example,
\begin{eqnarray}
&&\^{\ga_a}=\Pi_a^m\Ga_m \nonum \\
&&\^{h}_{ab}=\eta_{mn}\Pi^m_a\Pi^n_b
\end{eqnarray}  
The matrix  $\ga$ in (\ref{gamma}) is  written in a $D=10$ notation as
\begin{eqnarray}
\ga&=&\^{\ga}+\half \Ga_{11} \Pi^{11}_a\ep^{abc}\^{\ga}_{bc} \nonum \\
&=&(1-\^{\rho}\Ga_{11})\^{\ga} 
\end{eqnarray}
with
\begin{equation}
\^{\rho} = \Pi^{11}_a(\^{h}^{-1})^{ab}\^{\ga}_b
\end{equation}
We used the relation $\^{\ga}_a\^{\ga}=\half
\^{h}_{ab}\ep^{bcd}\^{\ga}_{cd}$. 
The matrix $\ga$ contains terms with both  even and odd
$\Ga^m$'s, and  it has no definite chirality-flip property in
10-dimensions. 
The ``irreducible'' $\ka$-symmetry is to choose 
$\ka^{ir}=\^{\ga}\ka_R /\^{\ga}^2$
and 
\begin{eqnarray}
\cob_{\ka}^{ir}\th &=& (1+\ga)\ka^{ir} \nonum \\
 &=&(1+\ga_{ir})\ka_R
\end{eqnarray}
where
\begin{equation}
\ga_{ir}=\^{\rho} +{\^{\ga} \over \^{\ga}^2}
\label{irr-gamma}
\end{equation}
$\cob_{\ka}^{ir}$  is essentially the $\ka$-symmetry of 
a $D2$-brane \cite{berg-town}. In the chiral 
notation, $\Ga_{11}\theta_{L,R} = \pm \theta_{L,R}$,  we have 
\begin{eqnarray}
\cob_{\ka}^{ir}\th_L &=& \ga_{ir}\ka_R \\
\cob_{\ka}^{ir}\th_R &=& \ka_R
\label{irr-ka-thR}
\end{eqnarray}
The matrix $\ga_{ir}$ contains  odd $\Ga_m$'s only, and consequently,   
$\ga_{ir}$ flips chirality in a 10-dimensional sense. The relation
$\cob_{\ka}^{ir}\th_R = \ka_R$ shows that the variable $\th_R$ is
identified with the gauge 
parameter $\kappa_{R}$ itself, which forms the basis of the $\th_R = 0$
gauge\cite{APS,berg-kall,kall}. Note that $\^{\ga} = 0$ for a naive
``double dimensional
reduction'',
and the gauge $\th_R = 0$ becomes singular in such a limit.  

By extending the covariant gauge fixing of the bosonic membrane, we adopt 
the gauge condition
\begin{equation}
\tg^{0a}+\cob^{0a}=0~,~~\th_R=0
\label{gauge-cond}
\end{equation}
The first condition
 corresponds to an orthogonal decomposition of the 
3-dimensional membrane world-volume $W$ into $W={\bf R}\times \Si$, where
${\bf R}$ and 
$\Si$ are time and space part of the membrane world-volume, respectively.
We use  $\tau\equiv \si^0$ and $\si^k~(k=1,2)$ for coordinates on
${\bf R}\times \Si$.
The gauge fixing and Faddeev-Popov terms are
\begin{eqnarray}
{\cal L}_g&=&\cob_{BRST}\bl[b_a(\t{g}^{0a}+\cob^{0a})
 +\b{\bt}_L\th_R \br] \nonum\\
&=&N_a(\t{g}^{0a}+\cob^{0a})+\b{\tilde{\xi}}_L\th_R \nonum \\
&&+ib_a[\del_b(c^b\tg^{0a})-\tg^{ba}\del_bc^0-\tg^{0b}\del_bc^a]
\label{FP-term}
\end{eqnarray}
with
\begin{equation}
\b{\tilde{\xi}}_L \equiv  \b{\xi}_L + i\del_a(\bar{\bt}_L c^{a})
\end{equation}
One may understand (\ref{FP-term}) as formally obtained
 from
\begin{eqnarray}
{\cal L}_g&=&N_a(\tilde{g}^{0a}+\cob^{0a})+\bar{\xi}_L\th_R \nonum \\
 &&+ ib_a\bl[\del_b(c^b\tilde{g}^{0a})-\tilde{g}^{ba}\del_b c^0
-\tilde{g}^{0b}\del_b c^a-K^a_b(\ka=\^{\ga}\ga_R/\^{\ga}^2)
\tilde{g}^{0b}\br] \nonum \\
&&+\bar{\bt}_L(\ga_R-ic^a\del_a\th_R)\nonum\\
&=&N_a(\tilde{g}^{0a}+\cob^{0a})+\bar{\tilde{\xi}}_L\th_R  \nonum\\
 &&+ ib_a\bl[\del_b(c^b\tilde{g}^{0a})-\tilde{g}^{ba}\del_b c^0
-\tilde{g}^{0b}\del_b c^a-K^a_b(\ka=\^{\ga}\ga_R/\^{\ga}^2)
\tilde{g}^{0b}\br] \nonum \\
&&+\bar{\bt}_L \ga_R
\end{eqnarray}
after partial integration  and then path integrating out the
{\em non-propagating}  ghost sector $\bar{\bt}_L \ga_R$; this 
procedure is analogous to that of the unitary gauge in the Higgs mechanism.
The variable $\xi_{L}$ is a Lagrangian multiplier to impose $\theta_{R}=0$,
 and $\gamma_{R}$ is a Faddeev-Popov ghost for $\kappa$-symmetry.
By this way , only the following reparametrization BRST symmetry remains in
 (\ref{FP-term})
\begin{eqnarray}
&&\cob X^{\mu}=-i\ep c^a\del_aX^{\mu} ~~,~~\cob\th =-i\ep c^a\del_a\th
\nonum \\
&&\cob c^a=-i\ep c^b\del_bc^a~~,~~\cob b_a=\ep N_a\nonum\\
&&\cob N_{a} = 0~~,~~\cob
\tilde{g}^{ab} = -i\ep [\del_c(c^c\tilde{g}^{ab})-\tilde{g}^{cb}\del_c c^a
-\tilde{g}^{ac}\del_c c^b]\nonum\\
&&\cob \bar{\tilde{\xi}}_L = -i\ep \del_a(c^a \bar{\tilde{\xi}}_L),
\end{eqnarray}
where the transformation law of $\bar{\tilde{\xi}}_L$ is induced by the
original transformation law of the Nakanishi-Lautrup multiplet
\begin{equation}
\cob \bar{\bt}_{L} = \ep \bar{\xi}_{L}~~,~~\cob \bar{\xi}_{L} = 0
\end{equation}
In some of the analysis of BRST symmetry\cite{fuji-ko2}, it is convenient
to revive an
extra
(redundant) variable $\bt_L$ by re-writing 
\begin{equation}
\b{\tilde{\xi}}_L\th_R \rightarrow \b{\xi}_L\th_R -
i\bar{\bt}_Lc^a\del_a\th_R
\end{equation}
In passing, we note that the choice of the gauge $\tilde{g}^{00} = - \rho
(\sigma^{1}, \sigma^{2})$ \cite{fuji-ko1} instead of $\tilde{g}^{00} = - 1$
in (\ref{gauge-cond})
introduces a density $\rho$ on $\Si$ described in \cite{dWHN}. In this
paper, we
work with the gauge choice (\ref{gauge-cond}).

\section{Gauge fixed action}
 
If we integrate out $\tg^{ab}$ and $N_a$ in the total action with the above
gauge
fixing Lagrangian (\ref{FP-term}), we obtain the  Lagrangian 
\begin{equation}
{\cal L}={\cal L}_0+{\cal L}_{WZ}+{\cal L}_{gh}
\label{th-non0-lag}
\end{equation}
where
\begin{eqnarray}
&&{\cal L}_0=\half(\Pi^\mu_0)^2 - \half\det G_{kl} \nonum \\
&&{\cal L}_{gh}=ib_0(\del_0c^0-{\rm div}\bfc)+i(\bfb,\del_0\bfc) +
\b{\tilde{\xi}}_L\th_R 
\end{eqnarray}
withXF
\begin{equation}
G_{kl}=\Pi^{\mu}_k\Pi_{\mu l}
 +ib_k\del_lc^0+ib_l\del_kc^0
\end{equation}
We defined the 1-form $\bfb$ and a vector field $\bfc$ on $\Si$ by 
\begin{equation}
\bfb=b_k\d \si^k~~,~~\bfc=c^k\del_k
\end{equation}
and $(\bfb,\del_0\bfc)$ stands for an inner product. 
The Wess-Zumino term ${\cal L}_{WZ}$, which is independent of the metric,
is not influenced in this procedure 
\begin{eqnarray}
{\cal L}_{WZ}&=&\Pi^{\mu}_0\{\Si_{\mu\nu},\d X^{\nu}+\half\Si^{\nu}\}\nonum
\\
&&-\half\Si^{\mu\nu}_0\lf(\{X_{\mu},X_{\nu}\}+\{\d X_{\mu}
+{1\o 3}\Si_{\mu},\Si_{\nu}\}\ri) \nonum \\
&&-\half\Si^{\mu}_0\{\Si_{\mu\nu},\d X^{\nu}+{1\o 3}\Si^{\nu}\}
\label{WZ-term}
\end{eqnarray}
We here defined ``Poisson bracket'' of two functions on $\Si$ by
\begin{equation}
\{f,g\}=\vep^{kl}\del_kf\del_lg=\del_1f\del_2g-\del_2f\del_1g
\end{equation}
where one regards the variables $(\sigma^{1}, \sigma^{2})$ on $\Si$ as
canonically  conjugate variables. In this paper, we use the  bracket
$\{f,g\}$ to denote the ``Poisson bracket'' in this sense; for the
conventional {\em true} Poisson bracket, we attach a suffix such as
$\{p,x\}_{P}$ to the bracket. 

We also defined the ``Poisson bracket'' of 1-forms
$A$ and $B$ on $\Si$ by 
\begin{equation}
\{A,B\}\equiv\ast (A \wg B) = \vep^{kl}A_kB_l=-\{B,A\}
\end{equation}
For two exact 1-forms, the ``Poisson bracket of 1-forms'' becomes  the
ordinary Poisson bracket of functions 
\begin{equation}
\{\d f,\d g\}=\{f,g\}
\end{equation}
For a general 1-form and an exact 1-form, we
have 
\begin{equation}
\{A,\d f\}=-(A,\vec{f}~)
\end{equation}
where $\vec{f}=\del_kf\vep^{kl}\del_l=\del_1f\del_2-\del_2f\del_1$ is a 
Hamiltonian vector field associated with $f$.

The reparametrization BRST symmetry now becomes 
\begin{equation}
\begin{array}{lcl}
\cob X^{\mu}=-i\ep c^a\del_aX^{\mu} &,&\cob \th=-i\ep c^a\del_a\th \\
\cob c^a=-i\ep c^b\del_bc^a &,& \cob b_a=\ep B_a \\
\cob\b{\tilde{\xi}}_L= -i\ep \del_a(c^a\b{\tilde{\xi}}_L)  &,& 
\end{array}
\end{equation}
where
\begin{eqnarray}
&&B_0=\half\Pi^{\mu}_0\Pi_{\mu 0}+\half\det G_{kl}
+2ib_0\del_0c^0+i\del_ab_0c^a+ib_k\del_0c^k \\
&&B_k=\Pi^{\mu}_0\Pi_{\mu k}+i\del_ab_kc^a+ib_k\del_0c^0+ib_l\del_kc^l
\end{eqnarray}
The variables $B_{a}$, up to equations of motion, correspond to the energy-
momentum tensor $T_{0a}$ on the world volume.
The BRST charge is given by
\begin{eqnarray}
Q_{BRST}&=&\int_{\Si} d^2\si\lf[c^0\bl(\half\Pi^{\mu}_0\Pi_{\mu 0}
+\half\det G_{kl}\br)+c^k\Pi^{\mu}_0\Pi_{\mu k}\ri.\nonum \\
&&~~~~~~~~~~~-ib_0(c^0\del_kc^k+c^k\del_kc^0)-ib_kc^l\del_lc^k
- c^0\b{\tilde{\xi}}_L\th_R ]
\end{eqnarray}
The above Lagrangian is regarded as a supersymmetrization of the bosonic
Lagrangian 
in Ref.\cite{fuji-kubo1}. 

The factor  $\det G_{kl}$ can be written as
\begin{equation}
\det G_{kl}=\half\{\Pi^{\mu},\Pi^{\nu}\}\{\Pi_{\mu},\Pi_{\nu}\}
+2i\{\Pi^{\mu},\bfb\}\{\Pi_{\mu},\d c^0\}-3(\bfb,\vec{c^0})^2
\label{det1}
\end{equation}
All the Poisson brackets of  1-forms, which appear in (\ref{WZ-term}) and
(\ref{det1}), are  reduced to the Poisson brackets of functions, for
example, 
\begin{eqnarray}
&&\{\Si^{\mu},\Si^{\nu}\}=\b{\th}\Ga^{\mu}\{\th,\b{\th}\}\Ga^{\nu}\th 
=\th^{\al}\Ga^{\mu}_{\al\bt}
\{\th^{\bt},\th^{\ga}\}\Ga^{\nu}_{\ga\cob}\th^{\cob} \\
&&\{\d X^{\mu},\Si^{\nu}\}=i\{X^{\mu},\b{\th}\}\Ga^{\nu}\th
\end{eqnarray}
We have thus established  an important fact: All the terms in the above
Lagrangian, which contain derivatives with respect to $\sigma^{k}$, except
for some of the ghost terms  are written in terms of the Poisson bracket of
functions. 

If we further integrate over $\b{\tilde{\xi}}_L$ and thus fix the gauge
$\th_R
= 0$ strictly, the Lagrangian is further simplified to
\begin{equation}
{\cal L}'={\cal L}'_0+{\cal L}'_{WZ}+{\cal L}'_{gh}
\label{th=0-lag}
\end{equation}
where
\begin{eqnarray}
{\cal L}'_0&=&\half(\Pi^m_0)^2+\half(\del_0 X^{11})^2-\half\det G_{kl}
\nonum \\
{\cal L}'_{WZ}&=&-\Si_m\d X^m\d X^{11} \nn
&=&i\b{\th}_L\Ga_m\bl[\del_0\th_L\{X^m,X^{11}\}
 +\del_0X^m\{X^{11},\th_L\}+\del_0X^{11}\{\th_L,X^m\}\br] \nonum \\
{\cal L}'_{gh}&=&ib_0(\del_0c^0-{\rm div}\bfc)+i(\bfb,\del_0\bfc)
\label{th=0-lag-rep}
\end{eqnarray}
with
\begin{eqnarray}
&&\Pi^m=\d X^m+\Si^m=\d X^m-i\b{\th}_L\Ga^m\d \th_L\\
&&G_{kl}=\Pi^m_k\Pi_{ml}+\del_kX^{11}\del_lX^{11}
 +ib_k\del_lc^0+ib_l\del_kc^0
\end{eqnarray}
The Wess-Zumino term  ${\cal L}'_{WZ}$ is  rewritten as
\begin{equation}
{\cal L}'_{WZ}=i\Pi^m_0 \b{\th}_L\Ga_m\{X^{11},\th_L\}
+i\del_0X^{11}\b{\th}_L\Ga_m\{\th_L,X^m\}-i\b{\th}_L\Ga^m\del_0\th_LY_m
\end{equation}
with
\begin{equation}
Y_m=-\{\Pi_m,\d X^{11}\}=-\{X_m,X^{11}\}+i\b{\th}_L\Ga_m\{\th_L,X^{11}\}
\label{def Y}
\end{equation}
The factor  $\det G_{kl}$ in (\ref{th=0-lag-rep}) is expanded as
\begin{eqnarray}
\det G_{kl}&=&\half\{\Pi^m,\Pi^n\}^2+\{\Pi^m,\d X^{11}\}^2 \nonum \\
&&+2i\{\Pi^m,\bfb\}\{\Pi_m,\d c^0\}+2i(\bfb,\vec{X^{11}})\{X^{11},c^0\}
-3(\bfb,\vec{c^0})^2 
\end{eqnarray}
The BRST symmetry is finally reduced to 
\begin{eqnarray}
&&\cob X^{\mu}=-i\ep c^a\del_aX^{\mu} ~~,~~\cob\th_L=-i\ep c^a\del_a\th_L
\nonum \\
&&\cob c^a=-i\ep c^b\del_bc^a~~,~~\cob b_a=\ep B_a
\end{eqnarray}
where
\begin{eqnarray}
&&B_0=\half\bl[(\Pi^m_0)^2+(\del_0X^{11})^2+\det G_{kl}\br]
+2ib_0\del_0c^0+i\del_ab_0c^a+ib_k\del_0c^k \\
&&\bfB=\Pi^m_0\Pi_m+\del_0X^{11}\d X^{11}+i\d (b_0c^0)+ib_0\d c^0
+i\bfb c-i{\cal L}_{\bfc}\bfb
\end{eqnarray}
with a Lie derivative of a 1-form $({\cal L}_{\bfc}\bfb)_{k} = c^l\del_l
b_k +
\del_kc^l b_l$ and $\bfB=B_k\d\si^k$. 
We also used the equations of motion, $\partial_{0}c^{0} = {\rm div }\bfc$
and 
 $\partial_{0}b_{k} = \partial_{k}b_{0}$, and we defined a new variable
\begin{equation}
c\equiv {\rm div }\bfc =\del_kc^k
\end{equation}
The BRST charge is then written as
\begin{eqnarray}
Q_{BRST}&=&\int_{\Si}\d^2\si\Br[\half c^0\bl(\Pi^m_0\Pi_{m0}
+(\del_0X^{11})^2+\det G_{kl}\br)
\nonum \\
&&+c^k(\Pi^m_0\Pi_{mk}+\del_0X^{11}\del_kX^{11})
-ib_0(c^0\del_kc^k+c^k\del_kc^0)-ib_kc^l\del_lc^k\Br]
\label{Q-BRST}
\end{eqnarray}
Lagrangians (\ref{th-non0-lag}) and (\ref{th=0-lag}) are physically
equivalent; Lagrangian (\ref{th-non0-lag}) exhibits more symmetry, whereas 
Lagrangian (\ref{th=0-lag}) contains the minimum set 
of variables in the present $\th_{R} = 0$ gauge.

\section{Superalgebra in path integral formulation}

We now examine the supersymmetry algebra in our formulation. We start with
the 
path integral analysis.  In this section, we use the action {\em before}
integrating out $N_a$ and $\tg^{ab}$:
\begin{equation}
S = S_{D} + S_{WZ} + \int_W\d^3\si  ~{\cal L}_{g}
\end{equation}
with the gauge fixing Lagrangian ${\cal L}_{g}$ in
(\ref{FP-term}).
The variation of the action under a localized SUSY is\cite{dWHN,rev-mem}
\begin{equation}
\cob_{\ep}S=\int_W \d^3\si\lf[i\del_a\b{\ep} J^a+\b{\ep}\tilde{\xi}_L
\ri]
+\int_W \half\d\bt_2
\label{susy-var-action}
\end{equation}
where,
\begin{eqnarray}
&& J^a=-2\t{g}^{ab}\Pi^{\mu}_b\Ga_{\mu}\th \nonum \\
&&~~~~~+\vep^{abc}\lf[\Ga_{\mu\nu}\th\Pi ^{\mu}_b\Pi ^{\nu}_c 
+{4i\o 3}(\Ga^{\mu}\th\b{\th}\Ga_{\mu\nu}\del_b\th
 +\Ga_{\mu\nu}\th\b{\th}\Ga^{\mu}\del_b\th)
(\Pi^{\nu}_c-{2\o 5}\Si^{\nu}_c) \ri]
\label{naive-supercharge}
\end{eqnarray}
and 
\begin{equation}
\bt_2 =l_{\mu\nu}\Bl[\al^{\mu\nu}_2 +2\Si^{\mu}
\lf(-{1 \over 3}\Pi^{\nu}+{1 \over 5}\Si^{\nu}\ri)\Br] \nonum \\
+l^{\mu}\Bl[{1 \over 3}\Si_{\mu\nu}\lf(\Pi^{\nu}-{4 \over
5}\Si^{\nu}\ri)\Br]
\label{beta-2}
\end{equation}
with $l^{\mu\nu}\equiv i\ep\Ga^{\mu\nu}\th$. In the evaluation of 
(\ref{naive-supercharge}) and (\ref{beta-2}),we used the identity
\begin{equation}
\Ga^{\mu\nu}_{(\al\bt}\Ga_{\mu\ga\cob)}=0
\end{equation}
where we symmetrize with respect to all the four spinor indices. 
The relation $\bar{\xi}\Gamma_{\mu}\eta = - \bar{\eta}\Gamma_{\mu}\xi$ for 
two Majorana spinors is also often used. 
The term  $\d\bt_2$ may  be
dropped for  a {\em closed} membrane, and we consider only this case in the
following. The equation of motion for the supercurrent is obtained from
(\ref{susy-var-action}) as
\begin{equation}
\del_a J^a=-i\tilde{\xi}_L
\label{eq-current}
\end{equation}
Physical information can be extracted from the equal-time commutator of
these
(broken-symmetry) supercharges in the path integral framework by using the
Bjorken-Johnson-Low(BJL) prescription \cite{fuji-ko2}. 

A SUSY transform of the supercurrent for a $\tau$-dependent but
$\si^k$-independent
parameter $\ep(\tau)$ is given by 
\begin{eqnarray}
-\cob_{\ep}(\b{\eta}J^0)&=&4\tg^{00}\del_0\b{\ep}\Ga^{\mu}\th\t{l}_{\mu}
+2\b{\ep}\Ga^{\mu}\eta P_{\mu}
-\b{\ep}\Ga^{\mu\nu}\eta \{X_{\mu},X_{\nu}\} \nonum \\
&&-\ast \d\lf[i(l_{\mu}\t{l}^{\mu\nu}+l^{\mu\nu}\t{l}_{\mu})
\bl({2\o 3}\d X_{\nu}+{1\o 5}\Si_{\nu}\br)
+{2i\o 15}(\t{l}^{\mu\nu}\Si_{\mu}+\t{l}_{\mu}\Si^{\mu\nu})l_{\nu}\ri]
\label{s-trf of J}
\end{eqnarray}
where the exterior derivative and the wedge operation are defined  on the
2-dimensional space $\Si$ of the membrane world-volume. We defined
$l^{\mu}=i\b{\ep}\Ga^{\mu}\th, \ \t{l}^{\mu}=i\b{\eta}
\Ga^{\mu}\th$ and the momentum density 
\begin{equation}
P^{\mu}={\cob {\cal L}\o \cob (\del_0X_{\mu})}
=-\tg^{0a}\Pi^{\mu}_a+\{\Si^{\mu\nu},\d X_{\nu}+\half\Si_{\nu}\} 
\end{equation}
The third term in (\ref{s-trf of J}) represents a (possible) central
charge of the
supermembrane. The fourth term 
in (\ref{s-trf of J}) is a total derivative , and one may think of it  as a
central charge also. However the fourth term  consists of terms containing
$\th$, and it may not give a non-vanishing contribution at the boundary.
 For this reason we tentatively drop it in the
following.

The Ward-Takahashi identity for SUSY generators  is given by
\begin{eqnarray}
&&{\del \o \del\tau } \<\Ts\b{\ep}Q(\tau)\b{\eta}Q(\t{\tau})\> 
\nonum \\
&=&4i{ \del \o \del\tau } \cob(\tau-\t{\tau})
\lf\< \int_{\Si} \d^2 \t{\si}~ 
\tg^{00}l^{\mu}\t{l}_{\mu}(\t{\tau},\t{\si}^k)\ri\> \nonum \\
&&+\cob(\tau-\t{\tau})\lf\<\int_{\Si} \d^2\t{\si}\bl[2\b{\ep}\Ga^{\mu}\eta 
P_{\mu}(\t{\tau},\t{\si}^k)
-\b{\ep}\Ga^{\mu\nu}\eta\{X_{\mu},X_{\nu}\}(\t{\tau},\t{\si}^k)\br]\ri\>
\nonum \\
&&+\lf\< -i\int_{\Si} \d^2\si ~\Ts\b{\ep}\xi_L(\tau,\si^k)\b{\eta}
Q(\t{\tau})\ri\>
\end{eqnarray}
In path integral this relation is obtained by starting with the expression 
\begin{equation}
\langle \b{\eta}Q(\t{\tau})\rangle
\end{equation}
and applying the localized SUSY  variation as in (\ref{s-trf of J}) and
also
the corresponding variation of the action.  The
above 
identity is written in terms of the $ \T^{\ast}$-product, and we can
identify 
the relation in terms of the $\T$-product by using the BJL prescription.
This 
procedure corresponds to removing the terms with
$\partial_{\tau}\cob(\tau-\t{\tau})$\cite{fuji-ko2}, and we obtain 
\begin{eqnarray}
&&{\del \o \del\tau } \<{\rm T}\b{\ep}Q(\tau)\b{\eta}Q(\t{\tau})\> 
\nonum \\
&=&\cob(\tau-\t{\tau})\lf\<\int_{\Si} \d^2\t{\si}\bl[2\b{\ep}\Ga^{\mu}\eta 
P_{\mu}(\t{\tau},\t{\si}^k)
-\b{\ep}\Ga^{\mu\nu}\eta\{X_{\mu},X_{\nu}\}(\t{\tau},\t{\si}^k)\br]\ri\>
\nonum \\
&&+\lf\<-i\int_{\Si} \d^2\si 
~\T\b{\ep}\xi_L(\tau,\si^k)\b{\eta}Q(\t{\tau})\ri\>
\end{eqnarray}
By explicitly operating the time derivative operation in  this relation and
using the equation of motion (\ref{eq-current}),
the {\em equal time}  commutator of supercharge is obtained as 
\begin{equation}
\bl[\b{\ep}Q(\tau),\b{\eta}Q(\tau)\br]
=\int_{\Si} \d^2\si~\b{\ep}\bl(2\Ga^{\mu}P_{\mu}(\tau,\si^k)
-\Ga^{\mu\nu}\{X_{\mu},X_{\nu}\}(\tau,\si^k)\br)\eta
\label{alg path-int}
\end{equation}
The gauge fixing term, which breaks  supersymmetry, does not influence
this equal-time SUSY algebra. This is analogous to the old chiral
$SU(3)\times
SU(3)$ algebra where the soft breaking of symmetry by a mass term does not
influence the chiral charge algebra. We here note that the derivation of 
(\ref{alg path-int}) goes through without modification for the Lagrangian
(\ref{th-non0-lag}) also.

If we compare the above algebra  with the most general $D=11$ super algebra
\begin{equation}
\{Q_{\al},Q_{\bt}\}=2\bl[P_{\mu}\Ga^{\mu}_{\al\bt}
+\half Z_{\mu_1\mu_2}\Ga^{\mu_1\mu_2}_{\al\bt}
+{1\o 5!}Z_{\mu_1..\mu_5}\Ga^{\mu_1..\mu_5}_{\al\bt}\br]
\label{d=11-alg}
\end{equation}
we can identify the central charge (density) of supermembrane as 
\begin{equation}
Z_{\mu\nu}=-\{X_{\mu},X_{\nu}\}
\end{equation}
Note that we obtain  no fivebrane charge $Z_{\mu_1\ldots\mu_5}$ in the 
superalgebra of supermembrane theory,
which  is consistent with the past light-cone gauge analyses of
superalgebra 
\cite{dWHN,matsuo1}.

In Matrix Theory as formulated in \cite{bfss}, one can in principle
introduce a longitudinal fivebrane  charge by using  two ``canonical
conjugate pairs of matrices''  \cite{banks-seiberg}.  Because of the
absence of  a transverse fivebrane charge  in the basic superalgebra of
matrix theory, the issue of  the 
Lorentz invariance of matrix theory becomes very subtle in the 
presence of the fivebrane. On the other hand, the superalgebra in 
supermembrane theory as evaluated here has a  manifestly Lorentz covariant
form.

In connection with the above path integral evaluation of superalgebra, we
want to comment on the following two issues: The first is if the path
integral itself is well-defined after $\theta_{R} = 0$ gauge fixing. The
second is a construction of a supercharge which is conserved, or if not
conserved, a supercharge which is conserved up to a BRST exact piece,
instead of the charge which is simply broken by the term
$\bar{\tilde{\xi}}_{L}\theta_{R}$ as in (\ref{eq-current}).

As is explained in the next section, we have a fermionic constraint 
$\chi_{\alpha}(\si) \approx 0$ (\ref{chi}) after the gauge fixing $\theta_{R} =
0$, and
the 
Poisson bracket of the constraint $\chi_{\alpha}(\si)$ is given by 
\begin{equation}
\{\chi_{\al}(\si),\chi_{\bt}(\t{\si})\}_P  = 
-2i\cob(\si-\t{\si})\Ga^m_{\al\bt}W_m(\si)
\end{equation}
with $W_m = P_m-\{X_m,X^{11}\}+2i\bar{\th}_L\Ga_m\{\th_L,X^{11}\}$. If one
treats this constraint as  a second-class constraint, one has to add an
extra term 
\begin{equation}
{\cal L}_{\phi_L}= \bar{\phi}_{L}\Ga^{m}W_{m}\phi_{L}
\label{L-phi}
\end{equation}
to the total Lagrangian  with a {\em bosonic} left-handed Majorana  spinor
$\phi_{L}$, and the path integral is defined by 
\begin{equation}
Z = \int d\mu{\cal D}\phi_{L}~ \exp\lf[ iS + i\int_W \d^{3}\sigma
\bar{\phi}_{L}\Ga^{m}W_{m}\phi_{L}\ri]
\end{equation}
This $\phi_{L}$ is analogous to the Faddeev-Popov ghost.\footnote[1]{
The origin of this factor is understood if one remembers that the
Faddeev-Popov gauge fixing is to add a constraint (gauge condition) 
$\chi_{2} \approx 0$ 
to make the first class gauge generator  $\chi_{1} \approx 0$  effectively
a second-class constraint. The Faddeev-Popov factor is then given by 
\begin{eqnarray}
\det \left[ \begin{array}{cc}
          \{\chi_{1},\chi_{1}\}_P & \{\chi_{1}, \chi_{2}\}_P \\
          \{\chi_{2},\chi_{1}\}_P & \{\chi_{2}, \chi_{2}\}_P \\
          \end{array}\right]^{1/2} = \det\{\chi_{1}, \chi_{2}\}_P \nonum
\end{eqnarray}
In the present case, we have the second class constraints $\chi_{\alpha}$,
and 
the determinant factor becomes $
(\det\{\chi_{\alpha},\chi_{\beta}\}_P)^{-1/2}$, where the minus sign in the
exponential arises from the fact that we are dealing with  fermionic
variables. This determinant factor is exponentiated by a bosonic spinor
$\phi_{L}$.} 
The field $\phi_{L}$ is non-propagating in the sense of a 3-dimensional
field theory, and it may be neglected if one applies a suitable
regularization.
However, this factor is important when one examines if the path integral is
well-defined. In fact, $(\Ga^{m}W_{m})^{2}\approx -(\del_0X^{11})^2 
-\half\{\Pi^m,\Pi^n\}^{2}$ as is explained in (\ref{W^2}), and it  vanishes
for
a naive vacuum configuration, which suggests that the path
integral could be singular even after the gauge fixing $\theta_{R} = 0$.
This singular behavior is presumably related to the well-known instability
of a supermembrane for a naive ground state configuration\cite{dWLN}. One
can of course stabilize the membrane by  a suitable compactification. The
non-trivial central charge generally suggests certain compactification, and
in this sense  our evaluation of the possible central charge is
well-defined.  Moreover , one can make the factor $\Ga^{m}W_{m}$ invariant
even under localized SUSY transformation if one uses a variable $p^{m} =
P^{m} + i\bar{\theta}_{L}\Gamma^{m}\{\theta_{L}, X^{11}\}$
 in a first order formalism, which means 
that ${\cal L}_{\phi_L}$ in (\ref{L-phi}) does not affect the analysis of
superalgebra.   

We next comment on the supercharge which is conserved up to a BRST exact
term. 
For this purpose, we rewrite the conservation equation (\ref{eq-current}).
The equation
of motion for $\theta_{R}$ gives
\begin{equation}
\bar{\tilde{\xi}}_{L}(\sigma) + \frac{\delta S^{(0)}}{\delta
\theta_{R}(\sigma)} = 0
\end{equation}
where we separated the total action $S = S^{(0)} + S_{g}$ into a gauge
fixing part $S_{g}$ and the rest. In the remainder of this section, the
derivative stands for the right-derivative. Next we note the
$\kappa$-symmetry of $S^{(0)}$,
which amounts to 
\begin{equation}
\int_W \d^{3}\sigma \lf[ \frac{\delta
S^{(0)}}{\delta\theta_{R}(\sigma)}\delta_{\kappa}
\theta_{R}(\sigma) + \frac{\delta
S^{(0)}}{\delta\theta_{L}(\sigma)}\delta_{\kappa}
\theta_{L}(\sigma) + \frac{\delta S^{(0)}}{\delta
X^{\mu}(\sigma)}\delta_{\kappa}X^{\mu}(\sigma) + \frac{\delta
S^{(0)}}{\delta \tilde{g}^{ab}(\sigma)}
\delta_{\kappa}\tilde{g}^{ab}(\sigma)\ri] = 0
\end{equation}
where we use the transformation law in (\ref{kappa-trans}). 
Since $S_{g}$ does not
depend on $X^{\mu}$ and $\theta_{L}$, the equations of motion give
$\frac{\delta S^{(0)}}{\delta\theta_{L}(\sigma)} =  \frac{\delta
S^{(0)}}{\delta X^{\mu}(\sigma)}=0$. The equation of motion for
$\tilde{g}^{ab}$ is given by
\begin{equation}
\frac{\delta S^{(0)}}{\delta \tilde{g}^{ab}(\sigma)} + \frac{\delta
S_{g}}{\delta \tilde{g}^{ab}(\sigma)} = 0
\end{equation}
If one combines the above 3 equations, one obtains
\begin{equation}
\bar{\tilde{\xi}}_{L}(\sigma) = - \frac{\delta S_{g}}{\delta
\tilde{g}^{ab}(\sigma)}\b{\psi}^{a}_{c}(\sigma)\tilde{g}^{cb}(\sigma)
\end{equation}
where we defined
\begin{equation}
K^{a}_{b}(\sigma) \equiv \b{\psi}^{a}_{b}(\sigma)\kappa_{R}(\sigma)
\end{equation}
in eq.(\ref{kappa-trans}) by noting (\ref{irr-ka-thR}).

For the specific gauge fixing Lagrangian (\ref{FP-term}), we obtain
\begin{eqnarray}
-\bar{\tilde{\xi}}_{L}&=&
N_{a}\b{\psi}^{0}_c\tilde{g}^{ca} -i
(\partial_{b}b_{a})c^{b}\b{\psi}^{0}_c\tilde{g}^{ca} 
-ib_{a}\bl[\partial_{b}c^{0}\b{\psi}^{b}_{c}\tilde{g}^{ca} +
\partial_{b}c^{a}\b{\psi}^{0}_{c}\tilde{g}^{cb}\br]
\nonum\\
&=&N_{a}\b{\psi}^0_{c}\tilde{g}^{ca} +ib_{a}\bl[
\partial_{b}(c^{b}\b{\psi}^{0}_{c}\tilde{g}^{ca})
 -\partial_{b}c^{0}\b{\psi}^{b}_{c}\tilde{g}^{ca} 
-\partial_{b}c^{a}\b{\psi}^{0}_{c}\tilde{g}^{cb}\br] \nn
&& -i\partial_{b}\bl(b_{a}
c^{b}\b{\psi}^0_{c}\tilde{g}^{ca}\br)\nonum\\
&=&
\delta_{BRST}\bl(b_{a}\b{\psi}^{0}_{c}\tilde{g}^{ca}\br)
-i\partial_{b}\bl(b_{a}
c^{b}\b{\psi}^{0}_{c}\tilde{g}^{ca}\br)
\end{eqnarray}
In the last equation, we used the fact that the transformation property of 
$\b{\psi}^{a}_{c}\ka_R\tilde{g}^{cb}$
under reparametrization symmetry is the same as $\tilde{g}^{ab}$,
since $\kappa$-symmetry does not interfere with the reparametrization
symmetry. As $\kappa_{R}$ is a scalar quantity under
reparametrization symmetry,
$\b{\psi}^{a}_{c}\tilde{g}^{cb}$ also has the same
transformation property as  $\tilde{g}^{ab}$.

We can thus rewrite eq.(\ref{eq-current}) as
\begin{equation}
\partial_{a}\tilde{J}^{a}=
i\delta_{BRST}\bl(b_{a}\psi^0_c\tg^{ca}\br)
\end{equation}
with
\begin{equation}
 \tilde{J}^{a}\equiv J^{a} - b_{b}c^{a}\psi^{0}_{c}\tilde{g}^{cb}
\end{equation}
One can also understand (\ref{susy-var-action}) up to equations of motion
as
\begin{equation}
\cob_{\ep}S=\int_W \d^3\sigma \bl(i\del_a\b{\ep} \tilde{J}^a
-\b{\ep}\delta_{BRST}(b_a\psi^0_{c}\tg^{ca})\br)
+\int_W \half\d\bt_2
\end{equation}
By starting with 
\begin{equation}
\langle \bar{\eta}\tilde{Q}(\tilde{\tau})\rangle
\end{equation}
where $\tilde{Q}(\tilde{\tau}) \equiv \int_{\Si} \d^{2}\sigma
\tilde{J}^{0}(\tilde{\tau}, \sigma^{k})$,  one can derive the
Ward-Takahashi identity as before, and one obtains the same algebra as
(\ref{alg path-int})
\begin{equation}
\bl[\b{\ep}\tilde{Q}(\tau),\b{\eta}\tilde{Q}(\tau)\br]
=\int_{\Si} \d^2\si~\b{\ep}\bl(2\Ga^{\mu}P_{\mu}(\tau,\si^k)
-\Ga^{\mu\nu}\{X_{\mu},X_{\nu}\}(\tau,\si^k)\br)\eta
\end{equation}
In deriving this relation, it is important to recognize that
$\psi^{0}_{c}(\sigma)$ depends on $\theta (\sigma)$ only through its
derivative $\partial_{a}\theta (\sigma)$ or SUSY invariant combination
$\Pi^{\mu}_{a}$. This means that the variation of the extra term in
$\tilde{J}^{a}$ under a $\tau$-dependent but $\sigma^{k}$-independent
supersymmetry transformation $\theta\rightarrow \theta + \ep (\tau)$ always
gives rise to terms proportional to $\partial_{\tau}\ep (\tau)$. These
terms
in turn give rise to terms proportional to 
$\partial_{\tau}\delta (\tau - \tilde{\tau})$, which are removed when one
goes to the $\T$-product from  $\T^{\ast}$-product by BJL
prescription\cite{fuji-ko2}.
Consequently,
the extra term in $\tilde{J}^{a}$ does not modify the supercharge algebra.

The supercharge $\tilde{Q}$ defined in terms of $\tilde{J}^{a}$ is
analogous to the conserved supercharge in the light-cone
gauge\cite{dWHN,rev-mem} which is
obtained by a suitable combination of the localized SUSY transformation and
$\kappa$-transformation.  Since the reparametrization gauge fixing of
$\tilde{g}^{0a}$ partly spoils
$\kappa$-symmetry, a choice of $\kappa_{R}$, which compensates the
variation of $\theta_{R}$ under a localized SUSY transformation $\ep_{R}$,
does not generate a conserved charge; instead one obtains $\tilde{Q}$
defined  above.

\section{Superalgebra via Dirac bracket}

 In this section, we present a Dirac bracket analysis of superalgebra 
on the basis of the Lagrangian (\ref{th=0-lag}), which directly leads to 
the commutator algebra in quantized theory. 
The conjugate momenta of $X^m,X^{11}$ and $\th_L$ are calculated from
Lagrangian (\ref{th=0-lag}) as 
\begin{eqnarray}
&&P_m=\Pi_{m0}+i\b{\th}_L\Ga_m\{X^{11},\th_L\} 
\label{P-m} \\
&&P_{11}=\del_0X_{11}+i\b{\th}_L\Ga_m\{\th_L,X^m\} \\
&&\b{S}_R=-i\b{\th}_L\Ga^m(P_m+Y_m)
\label{S-R}
\end{eqnarray}
where $Y_{m}$ is defined in (\ref{def Y}).
From the definition of (\ref{P-m}) and (\ref{S-R}),
we have a  primary constraint
\begin{equation}
\b{\chi}_R=\b{S}_R+i\b{\th}_L\Ga^m(P_m+Y_m) \approx 0
\label{chi}
\end{equation}
The constraint $\b{\chi}_R$ forms a  second class constraint (and
consequently,  there is  no secondary constraint). Using the standard
Poisson bracket relations
\begin{eqnarray}
\{S_{R\al}(\si),\th^{\bt}_L(\t{\si})\}_P &=& 
\cob_{\al}^{\bt}\cob(\si-\t{\si}) \nonumber\\
\{X^m(\si),P^n(\t{\si})\}_P &=& \eta^{mn}\cob(\si-\t{\si})
\end{eqnarray}
the Poisson bracket of the constraint $\chi_{\al}$ is calculated as
\begin{eqnarray}
C_{\al\bt}(\si-\t{\si}) &\equiv& 
 \{\chi_{\al}(\si),\chi_{\bt}(\t{\si})\}_P \nonum \\
&=& -2i\cob(\si-\t{\si})\Ga^m_{\al\bt}W_m(\si)  \nonum\\
(C^{-1})^{\al}_{~\t{\bt}}(\si-\t{\si}) &=& {i\over
2}\cob(\si-\t{\si}){\Ga^{m\al}_{~~~\t{\bt}}
W_m\over W^2}
\end{eqnarray}
where we defined 
\begin{equation}
W_m = P_m-\{X_m,X^{11}\}+2i\bar{\th}_L\Ga_m\{\th_L,X^{11}\}
=\Pi_{m0}+Y_m
\end{equation}
The Dirac bracket is generally defined as
\begin{equation}
\{f,g\}_D = \{f,g\}_P-\int\d^2\si\d^2\t{\si}\{f,\chi_{\al}(\si)\}_P
(C^{-1})^{\al}_{~\t{\bt}}(\si-\t{\si})\{\chi^{\t{\bt}}(\t{\si}),g\}_P
\label{dirac-bra}
\end{equation}
Supersymmetry  with a parameter $\ep_L$ is manifest in the $\th_R=0$ gauge.
The corresponding supercharge is obtained by the Noether procedure as 
\begin{equation}
Q_R =\int_{\Si} \d^2\si \;2\Ga_m\th_L\lf[\Pi^m_0-\{X^m,X^{11}\}+{1\o 3}
\{\Si^m,\d X^{11}\}\ri] 
\label{Q-R}
\end{equation}
For a supersymmetry transformation  with a parameter $\ep_R$, it is broken
by 
the gauge fixing term. We can nevertheless identify the supercharge $Q_L$
for 
the broken symmetry by simply substituting  the gauge conditions
$\tg^{0a}+\cob^{0a}=0$ and $\th_R=0$ into the  supercharge
(\ref{naive-supercharge}) defined from the original Lagrangian
\begin{eqnarray}
Q_L&=& \int_{\Si}\d^2\si \Bl(2P^{11}\th_L-\Ga^{mn}\th_L\{X_m,X_n\} \nonum
\\
&&+{2\o 3}\th_L\{\Si_m,\d X^m\}-{2\o 3}\Ga^{mn}\{\Si_m,\d X_n\}
-{1\o 5}\Ga^{mn}\th_L\{\Si_m,\Si_n\}\Br)
\label{Q-L}
\end{eqnarray}

We can establish  that $Q_L$ and $Q_R$ form a superalgebra in terms of the 
Dirac bracket. Using the data of Poisson brackets, for example, 
\begin{eqnarray}
&&\{Q^R_{\al},Q^R_{\bt}\}_P =-4\Ga^m _{\al\bt} 
\int_{\Si} \d^2\si\bar{\th}_L\Ga_m\{\th_L,X^{11}\} \nonum\\
&&\{Q^R_{\al},\chi_{\bt}\}_P=-2\Ga^m_{\al\bt}W_m  \nonum\\
&&\{Q_L^{\al},\chi_{\bt}\}_P=2\cob^{\al}_{~\bt}\del_0X^{11}-
\Ga_{mn\bt}^{~~\al}\{\Pi^m,\Pi^n\}
\label{bra-with-constr}
\end{eqnarray}
we can calculate the Dirac bracket of supercharges
\begin{eqnarray}
[ \b{\ep}_LQ_R,\b{\eta}_LQ_R ]_D &=& 2\b{\ep}_L\Ga^m\eta_L
 \int_{\Si}\d^2\si\bl(P_m-\{X_m,X_{11}\}\br) 
\label{dirac-alg1}  \\
\lf[ \b{\ep}_LQ_R,\b{\eta}_RQ_L \ri]_D &=& 2\int_{\Si}\d^2\si\bl(P^{11}
\b{\ep}_L\Ga_{11}\eta_R-\half\b{\ep}_L\Ga^{mn}\eta_R\{X_m,X_n\}\br) \nonum
\\
&&+\int_{\Si}\d\bl[i(\t{l}^{mn}l_m-\t{l}^{11}l^n)(2\d X_n+\Si_n)
+{4i\o 3}\t{l}^{11}l^n\Si_n\br]
\label{dirac-alg2}
\end{eqnarray}
where we defined $[~,~]_D\equiv i\{~,~\}_D$ ,$l^m=i\b{\ep}_L\Ga^m\th_L,
\t{l}^{11}=i\b{\eta}_R\Ga^{11}\th_L$ and
$\t{l}^{mn}=i\b{\eta}_R\Ga^{mn}\th_L$.  The last term in
(\ref{dirac-alg2}),
which depends on $\theta$, is neglected in the following. These algebraic 
relations are also readily derived by the path integral method by starting
with $\langle \bar{\eta}_{L}Q_{R}\rangle$ or $\langle
\bar{\eta}_{R}Q_{L}\rangle$ and using the BJL procedure in Lagrangian
(\ref{th=0-lag}).

We compare these Dirac brackets  with the chiral decomposition of
(\ref{d=11-alg}),
\begin{eqnarray}
\{Q^R_{\al},Q^R_{\bt}\}&=&2\bl[(P_m+Z_{m11})\Ga^m_{\al\bt}
+{1\o 5!}Z_{m_1..m_5}\Ga^{m_1..m_5}_{\al\bt}\br] \nonum\\
\{Q^R_{\al},Q^L_{\t{\bt}}\}&=&2\bl[P^{11}C_{\al\t{\bt}}
+\half Z_{mn}\Ga^{mn}_{\al\t{\bt}} 
-{1\o 4!}Z_{m_1..m_4 11}\Ga^{m_1..m_4}_{\al\t{\bt}}\br] \nonum\\
\{Q^L_{\t{\al}},Q^L_{\t{\bt}}\}&=&2\bl[(P_m-Z_{m11})
\Ga^m_{\t{\al}\t{\bt}}+{1\o
5!}Z_{m_1..m_5}\Ga^{m_1..m_5}_{\t{\al}\t{\bt}}\br]
\label{d=10-alg}
\end{eqnarray}
The central charge density can be read off from 
(\ref{dirac-alg1}) and (\ref{dirac-alg2}) as ,
\begin{equation}
Z_{m11}=-\{X_m,X_{11}\}~~,~~Z_{mn}=-\{X_m,X_n\}
\end{equation}
 
It is crucial to observe here  that we can reproduce the full algebra from
the commutators of 
$\{Q^R_{\al},Q^R_{\bt}\}$ and $\{Q^R_{\al},Q^L_{\t{\bt}}\}$ without using
the 
algebra $\{Q^L_{\t{\al}},Q^L_{\t{\bt}}\}$ whose evaluation is involved in
the  present strictly $\th_R = 0 $ gauge, as is explained below.

In the light-cone gauge, the kinetic terms of $X^{\mu}$ and $\th$ are
disentangled and they have the standard form\cite{dWHN}.So the Dirac
bracket in the
light-cone
gauge is very simple. In our $SO(9,1)$ invariant formulation, the Dirac
bracket
 is more involved. Some of the representative Dirac brackets are given by 
\begin{eqnarray}
&&\{\th_L^{\al}(\si),\th_L^{\bt}(\t{\si})\}_D=
  {i\o 2}{\Ws^{\al\bt}\o W^2}\cob(\si-\t{\si}) \nonum \\
&&\{X^m(\si),X^n(\t{\si})\}_D={i\o 2}{\b{\th}_L\Ga^m\Ws\Ga^n\th_L \o W^2}
 \cob(\si-\t{\si}) \nonum \\
&&\{X^m(\si),P^n(\t{\si})\}_D=\eta^{mn}\cob(\si-\t{\si})
  -{i\o 2}{\b{\th}_L\Ga^m\Ws\Ga^n\th_L \o W^2}
 \{X^{11}(\si),\cob(\si-\t{\si})\}
\end{eqnarray}
where  $\Ws=W_m\Ga^m$.
       
For the sake of completeness, we here present a Dirac bracket analysis of 
the algebra $[\b{\ep}_RQ_L,\b{\eta}_RQ_L]_D$. For this purpose, it turned
out 
to be simpler to work with the gauge $\theta_{R}=0$ but without the gauge
fixing of reparametrization symmetry, which avoids an analysis of the ghost
sector.
We thus start with the Nambu-Goto-type Lagrangian 
\begin{eqnarray}
&&{\cal L}={\cal L}_{NG}+{\cal L}_{WZ} \nonum \\
&&{\cal L}_{NG}=-\sqrt{-h} \nonum\\
&&{\cal L}_{WZ}=i\b{\th}_L\Ga_m\bl[\del_0\th_L\{X^m,X^{11}\}
 +\del_0X^m\{X^{11},\th_L\}+\del_0X^{11}\{\th_L,X^m\}\br] \nonum\\
&&h_{ab}=\Pi^m_a\Pi_{mb}+\del_aX^{11}\del_bX^{11}
\end{eqnarray}
which is obtained if we integrate out $\tg^{ab}$ and set $\th_R=0$ in 
(\ref{Dirac term}) and (\ref{Wess-Zumino}).
The conjugate momenta are defined by 
\begin{eqnarray}
&&P_m=p_m+i\b{\th}_L\Ga_m\{X^{11},\th_L\} \nonum\\
&&P_{11}=p_{11}+i\b{\th}_L\Ga_m\{\th_L,X^m\}\nonum\\
&&\b{S}_R=-i\b{\th}_L\Ga^m(P_m+Y_m)
\end{eqnarray}
where $p_m=-\sqrt{-h}(h^{-1})^{0a}\Pi_{ma},\ \  
p_{11}=-\sqrt{-h}(h^{-1})^{0a}\del_aX^{11}$  and  $Y_m=-\{\Pi_m,\d
X^{11}\}$.

There are second-class constraints
\begin{equation}
\b{\chi}_R=\b{S}_R+i\b{\th}_L\Ga^m(P_m+Y_m) \approx 0
\end{equation}
and first-class constraints, which correspond to the generators of
reparametrization symmetry, 
\begin{eqnarray}
&&B_0'=\half\lf(p_m^2+p_{11}^2+\det h_{kl}\ri)  \approx 0 \nonum\\
&&B_k'=p_m\Pi^m_k+p_{11}\del_kX^{11}\approx 0 
\label{B-prime}
\end{eqnarray}
The Poisson bracket of second-class constraint is given by
\begin{equation}
\{\chi_{\al}(\si),\chi_{\bt}(\t{\si})\}_P=-2i\cob(\si-\t{\si})\Ga_{\al\bt}^m
W_m
\end{equation}
with  $W_m=p_m+Y_m$.

The supercharge has the same form as in (\ref{Q-R}) and (\ref{Q-L})
,if we replace $\Pi^m_0$ and $\del_0 X^{11}$ by $p^m$ and $p^{11}$
,respectively. The Dirac bracket 
$[\b{\ep}_RQ_L,\b{\eta}_RQ_L]_D$ is given by (see (\ref{dirac-bra}) and
(\ref{bra-with-constr}))
\begin{eqnarray}
&&[\b{\ep}_RQ_L,\b{\eta}_RQ_L]_D=\int_{\Si}\d^2\si
{2\o W^2}\b{\ep}_R(p_{11}-Z)\Ws (-p_{11}-Z)\eta_R \nn
&&~~~~~~=\int_{\Si}\d^2\si{2\o W^2}\b{\ep}_R\lf\{(-p_{11}^2+Z^2)
\Ws+[Z,\Ws]p_{11}+\half [[Z,\Ws],Z]\ri\}\eta_R
\end{eqnarray}
where $Z=\half \Ga^{mn}\{\Pi_m,\Pi_n\}$ and $Z^2=-\half\{\Pi_m,\Pi_n\}^2$. 
The commutator of $\Ga$-matrices can be evaluated by noting 
$[\Ga^m,\Ga^{nl}]=2(\eta^{mn}\Ga^l-\eta^{ml}\Ga^n)$ as 
\begin{eqnarray}
 [Z,\Ws]p_{11}+\half [[Z,\Ws],Z]   
 &= &-2 W^m\{ \Pi_m,\Pi_n \}\bl(p_{11}\eta^{nl}+\{\Pi^n,\Pi^l\}\br)\Ga_l
\nn
 & \approx &-2\bl(-p_{11}Y_n+Y^m\{\Pi_m,\Pi_n\}\br)
\bl(p_{11}\eta^{nl}+\{\Pi^n,\Pi^l\}\br)\Ga_l \nn
&=&-2(-p_{11}^2+Z^2)Y_m\Ga^m
\label{keisan}
\end{eqnarray}
We used the relation
\begin{equation}
p^m\{\Pi_m,\Pi_n\}=-p_{11}Y_n+\{\bfB',\Pi_n\}\approx -p_{11}Y_n
\end{equation}
in the second step in (\ref{keisan}) by noting the constraints
(\ref{B-prime}).
$[\b{\ep}_RQ_L,\b{\eta}_RQ_L]_D$ is 
then reduced to
\begin{equation}
{[}\b{\ep}_RQ_L,\b{\eta}_RQ_L]_D=2\int_{\Si}\d^2\si{-p_{11}^2+Z^2 \o W^2}
(W_m-2Y_m)\b{\ep}_R\Ga^m\eta_R
\end{equation}
If we use 
\begin{eqnarray}
W^2&=&-p_{11}^2+Z^2+2B'_0-2\{\bfB',\d X^{11}\}
\nonum \\ &\approx&-p_{11}^2+Z^2 \nn
&=& -p_{11}^2  -\half\{\Pi_m,\Pi_n\}^2
\label{W^2}
\end{eqnarray}
by noting (\ref{B-prime}), we obtain the final result
\begin{eqnarray}
 [\b{\ep}_RQ_L,\b{\eta}_RQ_L]_D&\approx &2\b{\ep}_R\Ga^m\eta_R
\int_{\Si}\d^2\si (W_m-2Y_m) \nonum \\
&=&2\b{\ep}_R\Ga^m\eta_R\int_{\Si}\d^2\si (P_m+\{X_m,X_{11}\}) \nn
&=&2\b{\ep}_R\Ga^m\eta_R\int_{\Si}\d^2\si (P_m-Z_{m11})
\end{eqnarray}
which has a  form expected from (\ref{dirac-alg1}) and (\ref{d=10-alg}).
 To make our calculation
well-defined, we have to satisfy $-W^2 > 0$ in (\ref{W^2}).  Our analysis
of 
$[\b{\ep}_RQ_L,\b{\eta}_RQ_L]_D$ is analogous to the Hamiltonian analysis
of superalgebra for D-branes in \cite{kall}.

\section{Area preserving symmetry}
Our action (\ref{th=0-lag}) after 
integrating out the fields $\tilde{g}^{ab}$, $N_a$ 
and $\bar{\tilde{\xi}}$ has a symmetry under a shift of $\bfc$ by a
(fermionic) time-independent 
Hamiltonian vector field $\vec{\phi}$;~$\cob \bfc= \vec{\phi}$. (This
property
also holds for the action in (\ref{th-non0-lag}), but we analyze only 
 (\ref{th=0-lag}) here. Also , one can shift $\bfc$ by a more 
general  vector field ${\bf u}$ with $ {\rm div}{\bf u} = 0$ )
The generator of this symmetry is given by\cite{fuji-ko1}
\begin{equation}
V=*\d \bfb=\del_1b_2-\del_2b_1
\end{equation}
The area-preserving diffeomorphism (APD) of a general dynamical variable
${\cal O}$ is given by a 
Lie derivative with respect to a (bosonic) time-independent Hamiltonian
vector field $\vec{w}$;
\begin{equation}
\cob{\cal O}=-{\cal L}_{\vec{w}}{\cal O}
\end{equation}
For example, ${\cal L}_{\vec{w}}X^{\mu} = \{ w, X^{\mu}\}$.
The generator of APD is defined by 
\begin{eqnarray}
L&=&\ast\d\bfB  \nonum \\
&=& \{P_m,X^m\}+\{P_{11},X^{11}\}+\{\b{S}_R,\th_L\} \nonum \\
&&+i\{b_0,c^0\}+i(\bfb,\vec{c})+2iVc+i(\d V,\bfc)
\label{APD_L}
\end{eqnarray}
where $P_m,P_{11}$ and $\b{S}_R$ are defined in (\ref{P-m})-(\ref{S-R}).
It is shown  that the two generators $V$ and $L$ form a BRST multiplet
\cite{fuji-kubo2}
\begin{equation}
 L = \delta_{BRST}V
\end{equation}
which is physically understood if one remembers that $L$ generates a
reparametrization with a parameter of the form $\vec{w}$. 

The symmetries generated by $V$ and $L$ are characterized by
time-independent
parameters and thus analogous to the residual symmetry in $A_{0} = 0$ gauge
for Yang-Mills theory. We now discuss how to promote the symmetries
generated 
by $V$ and $L$ to  time-dependent gauge symmetry.
To gauge the APD symmetry, we rewrite the Lagrangian (in a first order
formalism)  by introducing  new {\em independent} variables $ P_m,P_{11}$
as 
\begin{eqnarray}
\t{{\cal L}}&=&P_m(\Pi^m_0+i\b{\th}_L\Ga^m\{X^{11},\th_L\})
+P_{11}(\del_0X^{11}+i\b{\th}_L\Ga^m\{\th_L,X_m\}) 
-\half P_m^2-\half P_{11}^2 \nonum \\
&&+\half(\b{\th}_L\Ga^m\{X^{11},\th_L\})^2
+\half(\b{\th}_L\Ga^m\{\th_L,X_m\})^2 -\half\det G
-i\b{\th}_L\Ga^m\del_0\th_LY_m \nonum \\
&&+ib_0(\del_0c^0-{\rm div}\bfc)+i(\bfb,\del_0\bfc)
\label{phase-space-lag}
\end{eqnarray}
If we integrate out $ P_m$ and $P_{11}$, we go back to the original
Lagrangian
(\ref{th=0-lag}). The Lagrangian  (\ref{phase-space-lag}) has an 
APD symmetry if we
assign a transformation 
property to  $P_m$ and $P_{11}$ as functions on $\Si$. The APD generator
has the same form as (\ref{APD_L}), but $P_m$ and $P_{11}$ are now 
independent variables.

We now introduce a BRST doublet $A,\la$\cite{fuji-ko1}
\begin{equation}
\cob_{BRST}A=-i\la~~,~~\cob_{BRST}\la=0
\end{equation}
as gauge fields for $L$ and $V$, respectively.
If we add a BRST exact term
\begin{equation}
{\cal L}_{exact}=-\cob_{BRST}(AV)=i\la V-AL
\end{equation}
to the Lagrangian (\ref{phase-space-lag}), which does not change the
physical 
contents of the theory,  and if we integrate out
$P_m$ and $P_{11}$, the Lagrangian becomes
\begin{eqnarray}
{\cal L}_{gauged}&=&\half(D_0X^m-i\b{\th}_L\Ga^mD_0\th_L)^2
+\half(D_0X^{11})^2
-\half\det G_{kl} \nonum \\
&&+i\b{\th}_L\Ga_m\bl(D_0\th_L\{X^m,X^{11}\}+D_0X^m\{X^{11},\th_L\}
+D_0X^{11}\{\th_L,X^m\}\br) \nonum \\
&&+b_0(D_0c^0-{\rm div}\bfc)+i(\bfb,D_0\bfc+\vec{\la})
\label{gauged-lag}
\end{eqnarray}
where $D_0$ is an ``APD covariant derivative'' defined by 
\begin{equation}
D_0{\cal O}=\del_0{\cal O}+{\cal L}_{\vec{A}}{\cal O}
\end{equation}
The Lagrangian ${\cal L}_{gauged}$ has an area preserving gauge symmetry
\begin{eqnarray}
&&\cob_V\bfc=\vec{\phi}~~,~~\cob_V\la=-D_0\phi \nonum \\
&&\cob_L{\cal O}=-{\cal L}_{\vec{w}}{\cal O}~~,~~\cob_LA=-D_0w
\end{eqnarray}
for  {\em time dependent} functions $\phi (\tau,\si^k)$ and 
$w(\tau,\si^k)$.
Note that the Lagrangian (\ref{gauged-lag}) has a structure quite different
from that of the  M(atrix) theory
Lagrangian of Refs.\cite{dWHN}\cite{bfss}, and it is not  given by a simple
dimensional-reduction of $D=10$  super Yang-Mills theory.
 The Lagrangian (\ref{th=0-lag}) corresponds to the gauge 
fixing $A = \lambda = 0$ of
this area preserving gauge symmetry. 

\section{$SO(9,1)$ covariant matrix regularization}

The physical state conditions in our formulation (\ref{th=0-lag}) are given
by
\begin{eqnarray}
&&V|{\rm phys}\>=\d\bfb|{\rm phys}\>=0 \nonum\\
&&L|{\rm phys}\>=\d\bfB|{\rm phys}\>=0
\label{V=L=0}
\end{eqnarray}
,which are the Gauss-law constraints for the $A=\la=0$ gauge in 
(\ref{gauged-lag}),      
in addition to the BRST invariance $ Q_{BRST}|{\rm phys}\rangle = 0 $. We
now {\em locally} solve $V =  0$  at the operator level
\begin{equation}
\bfb=-\d b
\label{def of b}
\end{equation}
and treat $c={\rm div}\bfc$ as an independent variable. This procedure is
shown to correspond to a gauge fixing of the symmetry generated by $V$ by a
gauge condition $F = \del_1 c^2 - \del_2 c^1 = 0 $\cite{fuji-ko1}. 
The Lagrangian (\ref{th=0-lag})  is then written as 
\begin{eqnarray}
{\cal L}&=&\half(\Pi^m_0)^2+\half(\del_0X^{11})^2-\half\det G' 
+ib_0(\del_0c^0-c)+ib\del_0c\nonum \\
&&+i\b{\th}_L\Ga_m\bl[\del_0\th_L\{X^m,X^{11}\}
 +\del_0X^m\{X^{11},\th_L\}+\del_0X^{11}\{\th_L,X^m\}\br]
\label{v=0lag}
\end{eqnarray}
with
\begin{eqnarray}
\det G'&=&\half\{\Pi^m,\Pi^n\}^2+\{\Pi^m,\d X^{11}\}^2 \nonum \\
&&+2i\{\d b,\Pi^m\}\{\Pi_m,\d c^0\}+2i\{b,X^{11}\}\{X^{11},c^0\}
-3\{b,c^0\}^2
\end{eqnarray}
It is crucial that all the  terms in (\ref{v=0lag}) which contain
derivatives 
with respect to the variables $ (\sigma^1, \sigma^2)$ are written in terms
of 
the ``Poisson bracket of functions'' on $\Si$. We can thus 
 matrix-regularize the Lagrangian in a formal way  by the ``correspondence
principle''\cite{dWHN};
\begin{equation}
\begin{array}{ccc}
{\cal O}_A(\tau)Y^A(\si^1,\si^2)&\riya& {\cal O}_A(\tau)T^A \\
\int_{\Si} \d^2\si &\riya& \Tr \\
\{~,~\} &\riya& -i[~,~]
\end{array}
\end{equation}
for a generic dynamical variable ${\cal O}$; $\{Y^A(\si^1,\si^2)\}$ are a
complete set of  orthonormal
eigenfunctions 
of Laplacian on $\Si$, and $\{T^A\}$ are the generators of $SU(N)$  with 
$N \rightarrow \infty $.

The matrix-regularized action is then written as 
\begin{eqnarray}
S&=&\int\d \tau \Tr\Bl[\half(\Pi^m_0)^2+\half(\del_0X^{11})^2-\half\det G' 
+ib_0(\del_0c^0-c)+ib\del_0c \nonum \\
&&+\b{\th}_L\Ga_m\bl(\del_0\th_L[X^m,X^{11}]
 +\del_0X^m[X^{11},\th_L]+\del_0X^{11}[\th_L,X^m]\br)\Br] 
\end{eqnarray}
where
\begin{eqnarray}
\det G'&=&-\half\bl([X^m,X^n]-i\b{\th}_L\Ga^m[\th_L,X^n]
-i\b{\th}_L\Ga^n[X^m,\th_L]+\b{\th}_L\Ga^m[\th_L,\b{\th}_L]_+\Ga^n\th_L\br)^2
\nonum \\
&&-\bl( [X^m,X^{11}]-i\b{\th}_L\Ga^m[\th_L,X^{11}]\br)^2 \nonum \\
&&-2i\br([b,X^m]+i\b{\th}_L\Ga^m[b,\th_L]_+\br)
\bl([X_m,c^0]-i\b{\th}_L\Ga_m[\th_L,c^0]_+\br) \nonum \\
&&-2i[b,X^{11}][X^{11},c^0]+3[b,c^0]_+^2
\end{eqnarray}
The bracket $ [b,c^0]_{+} $, for example, stands for an anti-commutator of
matrix
valued fermionic variables.

The Hamiltonian and the generator of the area preserving diffeomorphism are
respectively represented in  matrix formulation as
\begin{equation}
H = \Tr \lf( \frac{1}{2} p^\mu p_\mu + \frac{1}{2}\det G' + ib_0 c\ri)
\label{hamiltonian}
\end{equation}
and 
\begin{equation}
\Tr (w L^{\prime}) = \Tr w \bl(-i[P_m, X^m ]  -i[P_{11}, X^{11}]
-i[\bar{S}_R,
\th_L]_
{+} + [b_0, c^0 ]_{+} +[b,c]_{+}\br) 
\end{equation}
where the parameter $w (\sigma^1, \sigma^2)$ is also represented by an
infinite
dimensional matrix. In the Hamiltonian above  we defined the variables 
\begin{eqnarray}
&&p_m \equiv P_m - \b{\th}_L\Ga_m [X^{11},\th_L] \nonum\\
&&p_{11}\equiv P_{11} - \b{\th}_L\Ga_m [\th_L,X^m] \nonum\\
&&\b{S}_R = -i\b{\th}_L\Ga^m(P_m +i[X_m,X^{11}] +\b{\th}_L\Ga_m
{[}\th_L,X^{11} {]} )
\label{small-p}
\end{eqnarray}
In the Hamiltonian formulation we have a ( second class) constraint
\begin{equation}
\chi = \b{S}_R  + i\b{\th}_L\Ga^m(P_m +i[X_m,X^{11}] +\b{\th}_L\Ga_m
{[}\th_L,X^{11}{]}) \approx 0
\end{equation}
which complicates  practical manipulations, although the Hamiltonian itself
has a relatively simple form as in (\ref{hamiltonian}).

The supercharge is defined in the chiral decomposition as 
\begin{equation}
Q_R =\Tr \Bl[2\Ga_m\th_L \bl(P^{m} - \bar{\th}_{L}\Ga^{m}[X^{11},\th_{L}] +
i
[X^m,X^{11}] - \frac{1}{3}[X^{11}, \bar{\th}_{L}]\Ga^{m}\th_{L}\br)\Br] 
\end{equation}
which corresponds to the Noether charge for $\delta\th_{L} = \ep_{L}$ and
$\delta X^{\mu} = i\bar{\ep}_{L}\Ga^{\mu}\th_{L}$, and 
\begin{eqnarray}
Q_L&=& \Tr \Bl( 2P^{11}\th_L + i\Ga^{mn}\th_L[X_m,X_n] 
 - {2\o 3}\th_L\b{\th}_L\Ga^m[\th_L,X_m]  \nonum \\
&&+ {2\o 3}\Ga^{mn}[X_{n},\bar{\th}_{L}]\Ga_{m}
\th_{L} + {i\o 5}\Ga^{mn}\th_L\bar{\th}_{L}\Ga_m[\th_{L},
\bar{\th}_{L}]_{+}
\Ga_{n}\th_{L}\Br)
\end{eqnarray}
which corresponds to $\delta\th_{R} = \ep_{R}$ and $\delta X^{\mu} =
i\bar{\ep}_{R}\Ga^{\mu}\th_{R}$ in (\ref{th-non0-lag});
 the variable $\th_{R}$ is set to
$0$ after the 
evaluation of the Noether charges.

We here note that the reparametrization BRST charge $Q_{BRST}$ in 
(\ref{Q-BRST})
itself does not have a simple matrix representation. For example, the
replacement of 
$c^{k}$ by a single  variable $c={\rm div}\bfc$ does not go through, and
the
two
variables $c^{1}$ and $c^{2}$ remain in the BRST charge till the end. A
true 
significance of this property is not clear, but it is partly related to the
fact that we solved $V=0$ in the operator level but $L=\delta_{BRST}V = 0$
is not solved in the operator level: Instead we impose a constraint
$L|{\rm phys}\rangle =0$ on the physical state vector. The manifest BRST
invariant formulation in (\ref{V=L=0}) is thus partly spoiled by
 solving $V=0$ in
the above procedure. 

The BRST charge $Q_{BRST}$ generates transformation with 3 independent
ghost
variables $(c^{0}, c^{1}, c^{2})$, whereas the matrix formulation above
contains only two independent ghost variables $(c^{0}, c={\rm div}\bfc)$
and corresponding canonical conjugate variables $(b_{0}, b)$. This
reduction of the number of freedom is related to the symmetries generated
by $V$ and $L$, which exist even after the 
BRST invariant reparametrization gauge fixing in (\ref{FP-term}). 
This suggests that
a proper use of the area preserving diffeomorphism reduces the BRST
invariant 
physical states to those specified by two independent ghosts $(c^{0},
c={\rm div}\bfc)$ only. Although we cannot implement this statement in the
operator level, we expect that this procedure works in the physical matrix 
elements formed by BRST invariant physical states.

\section{Discussion}

We have presented an $SO(9,1)$ invariant formulation of the 11-dimensional
supermembrane by combining an $SO(10,1)$ invariant treatment of
reparametrization symmetry with an $SO(9,1)$ invariant $\theta_{R} = 0$
gauge of  $\kappa$-symmetry. The light-cone gauge formulation\cite{dWHN} 
and the present $SO(9,1)$ invariant formulation, for example 
eq.(\ref{gauged-lag}), have quite different 
appearance. We however emphasize that these two formulations in fact
describe an {\em identical} theory (i.e., 11-dimensional supermembrane)
in the common domain where both gauge conditions are well-defined.

Our $SO(9,1)$ formulation of supermembrane compared to the light-cone gauge
formulation preserves a large subset of $D=11$
Lorentz boost  symmetry , which may be regarded as ``dynamical'' symmetry.
However,  the rotational symmetry between
``M-direction''($X^{11}$-direction) and the other directions is not
manifest. Recently $D=11$ Lorentz symmetry 
has been  checked in  a $D2$-brane scattering with M-momentum
transfer \cite{polchinski} and in the analysis of Lorentz
algebra in the light-cone gauge \cite{matsuo1}.
 A similar check need to be done in our formalism as to  the
Lorentz-algebra  
and also the explicit calculation of  dynamical processes.

It is known that Green-Schwarz type IIA string action is obtained by a 
``double dimensional reduction'' of the supermembrane action\cite{mem-2a}.
 The $\theta_{R}=0$ gauge 
becomes singular for a naive ``double dimensional reduction'' 
due to the denominator $\^{\ga}^2$ in  $\ga_{ir}$ (\ref{irr-gamma}),
 and in this sense
our  $SO(9,1)$ invariant formulation may be regarded as 
intrinsic to the supermembrane.  However,  $\th_R=0$ 
gauge is smoothly related to the configuration which  has no winding in
M-direction, i.e., the  $D2$-brane.

Recently the supermembrane with non-trivial winding has been  studied
\cite{dWPP,  matsuo2}.
In these analyses, harmonic
 1-forms on $\Si$ play an essential role.
In our treatment of $V$ and $L$ symmetries, we considered only  Hamiltonian
vector fields. In order to treat the non-trivial topology of $\Si$ and the 
supermembrane with winding, we have to consider area-preserving
diffeomorphism
(APD) corresponding  to a
``locally Hamiltonian vector field'', which is a symplectic dual of the
 harmonic 1-form. We have additional constraints associated with  ``locally
Hamiltonian vector fields'' ${\bf u}_i~(i=1,\cdots,2g)$ on $\Si$ of genus
$g$
\begin{eqnarray}
&&V_i=\int_{\Si}\d^2\si (\bfb,{\bf u}_i) \approx 0 \\
&&L_i=\int_{\Si}\d^2\si (\bfB,{\bf u}_i)=\cob_{BRST}V_i \approx 0 
\end{eqnarray}
which are the generators of symmetries $\cob_{V_i}\bfc=\ep  {\bf u}_i$
and $ \cob_{L_i}{\cal O}=-{\cal L}_{{\bf u}_i}{\cal O}$,respectively.
   When we solved the constraint $\d\bfb=0$ in (\ref{def of b}),
 we neglected the
harmonic part of $\bfb$, which is justified only when $\Si={\rm S}^2$. It
is generally necessary to 
consider the effect of harmonic 1-forms when we matrix-regularize the 
supermembrane with genus $g \geq 1$.

As to the relevance of our formulation to the so-called M-theory, a better 
understanding of  the fundamental degrees of freedom in M-theory is 
important. The  M(atrix) theory in \cite{bfss} is formulated by regarding 
$D0$-branes as fundamental degrees of freedom. A crucial observation  is
the 
decoupling of anti-$D0$-branes in the  infinite momentum frame, and it
allows 
them  to treat only $D0$-branes as fundamental degrees of freedom.
But in a general Lorentz frame, the  interaction between $D0$-branes and
anti-$D0$-branes cannot be ignored in general, and a deeper understanding
of  the fundamental degrees of freedom is required. 
The basic idea of ``membrane as  composites of $D0$-branes''by
Townsend\cite{townsend} , which is one of  the physical bases  of M(atrix)
theory, is based on the resemblance of the light-cone gauge action of
supermembrane (APD ${\rm SYM}_{0+1}$)\cite{dWHN} and the
effective action of $N$ coincident $D0$-branes ($U(N) ~{\rm SYM}_{0+1}$)
\cite{witten}. Our
$SO(9,1)$ invariant Lagrangian with APD gauge symmetry (\ref{gauged-lag})
does  not correspond to a dimensional reduction of $D=10$ supersymmetric
Yang-Mills theory, and a direct relation to $D0$-branes is lost.  However,
our 
Lagrangian and light-cone Lagrangian describe the identical physics as was 
emphasized above, and we hope that our formulation may shed new light on
the 
basic dynamics of $D0$-branes and anti-$D0$-branes,  and possibly on the 
dynamics of M-theory itself.


\end{document}